\documentclass[manuscript,nonacm]{acmart}

\setcopyright{rightsretained}
\copyrightyear{2021}

\acmConference[SimuRec]{Workshop on Simulation Methods for Recommender Systems}{Sep. 26--30}{Amsterdam and Online}

\begin{document}

\title{Multiversal Simulacra: Understanding Hypotheticals and Possible Worlds Through Simulation}

\author{Michael D. Ekstrand}
\email{michaelekstrand@boisestate.edu}
\orcid{0000-0003-2467-0108}
\affiliation{%
  \institution{People and Information Research Team, Boise State University}
  \city{Boise}
  \state{Idaho}
  \country{USA}
  \postcode{83725-2055}
}

\begin{abstract}
  Recommender systems research is concerned with many aspects of recommender system behavior and effects than simply its effectiveness, and simulation can be a powerful tool for uncovering these effects.
  In this brief position paper, I identify specific types of research that simulation is uniquely well-suited to address along with a hierarchy of simulation types.
\end{abstract}




\maketitle

\section{Introduction}

My research agenda is particularly concerned with understanding the \textbf{human biases} that affect information retrieval and recommender systems, and quantifying their impact on the system's operation, individual and social human experience, and our metrics for quantifying operation, behavior, and experience.
This agenda requires significant use of simulation for a variety of reasons, most stemming from the need to study counterfactuals.
We can train a recommender system and measure its behavior on a data set, or on multiple different data sets that differ in key ways, but with such ``found data'' we do not have the ability to isolate specific phenomena and train on data that differ only in the way we want to study (e.g. the degree of popularity bias or preferential attachment users exhibit, or discriminatory biases in how users select and engage with content).

Simulation and synthetic data each, in different ways, give us a powerful tool for isolating these phenomena and quantifying their (expected) impact.
They allow us to study how the system \textit{would} behave under alternative conditions, both to better understand the conditions of our present world and to anticipate the effects of social change (whether that change is endogenous or is cultivated through policy and/or education).
This work is not easy, as the simulation must be credible, and it loses a certain documented fidelity to world it in which recommender systems will actually be deployed, but in exchange it opens the doors to questions that are intractable otherwise.

I approach this from the understanding that the goal of scientific research on recommender systems is to understand how our world works, and how recommenders systems work or can be made to work in the world we inhabit.
However, there is much we do not know about the relationship of our observations to the world; our observations may, in fact, be compatible with many different ``true'' configurations of the world.
One approach, therefore, is to understand system behavior in multiple possible worlds, or hypothetical, worlds, in hopes that our actual world is represented among them and that additional research may narrow down the set that contains our world, and simulation can enable that.
This conception also has the benefit of providing tools to try to understand how recommendation would work in worlds that ours could transform into, for example if a particular educational effort to combat a kind of human bias were effective.

\section{Simulation and Data}

\begin{figure}
    \centering
    \includegraphics[width=\textwidth]{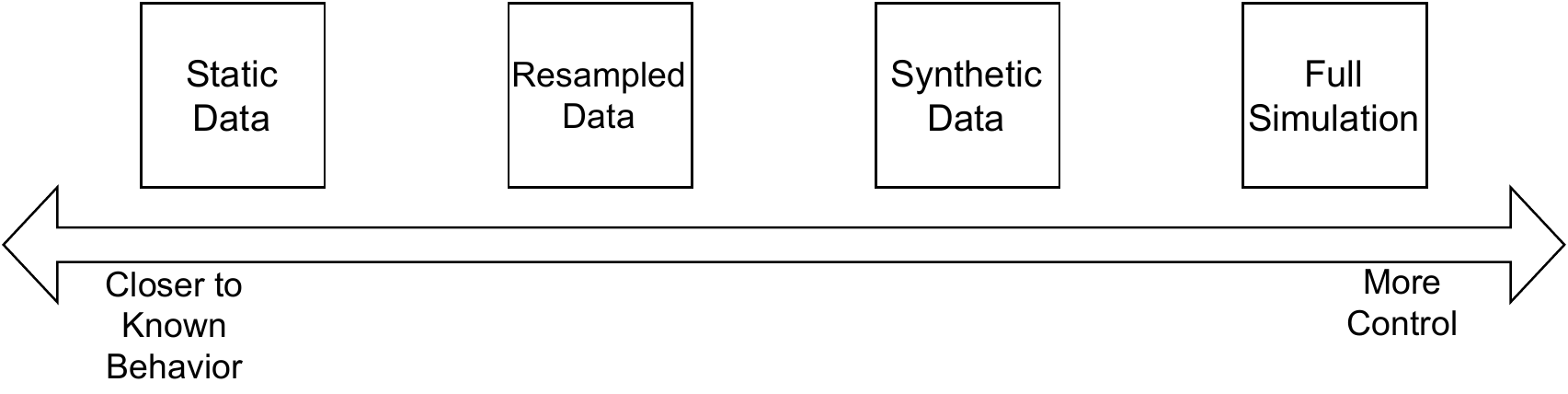}
    \caption{Spectrum of known partial fidelity and control.}
    \label{fig:spectrum}
\end{figure}

Before getting in to applications of simulation, I want to first lay out my perspective on the space of simulation methods and how different methods relate to each other.
These methods sit on a spectrum (see Fig.~\ref{fig:spectrum}), allowing an increasing degree of control over experimental conditions at the expense of decreased connection to observed behavior or interactions.

Much recommender systems research, including the vast majority of my own, uses existing \textbf{static data} to train recommendation models and evaluate their behavior or effectiveness.
These offline experiments can be viewed as a kind of simulation on their own \citep[][\S 2.5]{Ekstrand2021-bl}, because they use historical data traces to simulate what the system would do --- and how users would respond --- if it were actually applied to the task.
There are many limitations to this approach (\citet{Kouki2020-nv} provide one recent example of work exploring these limitations) but it is highly approachable and widely-used.
It has the advantage of being actual data collected from an actual system, along with the confounds that introduces \citep{Chaney2018-us} and a lack of knowledge about actual user characteristics and preferences behind the noisy observation process that yielded the recorded data.

A typical static data set is the result of a particular population interacting with a particular (suite of) recommendation algorithms in one system.
While other data sets reflect different populations, algorithms, domains, and behaviors, they often vary across multiple dimensions simultaneously, making it difficult to isolate which features of the data drive a behavior of interest.
Very little research has done this yet, but it seems possible to \textbf{resample data} to create a new data set that is as like an existing one as possible, except for altering the distribution with respect to a particular feature.  This can be done, for example, by reweighting data to ensure equal gender representation \citep[e.g.][]{Boratto2021-nv}.
This method increases our control from static data while maintaining a certain fidelity to the underlying collected data.

We can also generate \textbf{synthetic data} from a model that is trained to emulate existing data sets in key ways; this can be fully synthetic data \citep{Slokom2018-hn} or it can be synthetic attributes to go with existing data \citep{Burke2018-ua}.
There are a lot of open questions about how to ensure the data generator produces data that is realistic, as well as how to ensure the experimental process is truly studying recommender performance and not just the ability of the recommender to recover the data generator's parameters, but it gives us full control over the composition of a data set (or a family of data sets), and we can produce multiple data sets that truly differ in only one parameter.

Finally, we can run a \textbf{full simulation} that produces a data set, and then simulates user response and system behavior over repeated training or online learning.
This has been adopted for evaluating reinforcement learning agents \citep{Rohde2018-ej} as well as understanding system dynamics \citep{Chaney2018-us}, and has significant potential for helping the community more fully understand how recommender systems work in practice and their response to phenomena of concern.

\section{Retrospective Simulation: Studying Assumptions}

One of the major challenges to recommender systems research is the opacity of the data generating process and the various assumptions we must make about the relationship between observed data (whether in an offline, static data set or collected online) and underlying user preferences and needs.
\citet{Friedler2021-ns} describe the relationship of observations and underlying mechanisms as that of underlying ``construct spaces'' (the construct feature space, in which entities' true representations lie, and the construct decision space, representing ideal outcomes under complete and perfect information), that we observe through an observation process to obtain the ``observed feature space'' (how entities are represented after the incomplete and possibly biased observation process) and the ``observed decision space'' representing the decisions we make on the basis of these observed features.
Their work was focused on algorithmic fairness, and grounding many fairness concerns in \emph{distortions} between the various spaces; however, the framework for understanding data and decisions is far more general.
In recommender systems, we can think of the construct feature space as holding users' true preferences (or their time-varying and context-specific constituent components) and the construct decision space as representing their ideal recommendations; the observed feature space is the representation of users we can actually obtain through the data they make available to us.

There are a variety of distortions that can occur between the construct and observed feature spaces, that can systematically affect both the \emph{presence} of observations (recommender system data is missing-not-at-random \citep{Marlin2007-fo, Marlin2009-ch}) and the \emph{values} of those observations.
Some of these distortions are well-documented in the recommender systems literature, such as \emph{popularity bias} \citep{Canamares2018-ki}.

Two of the impacts of biased observation process are that the system \textbf{learns from biased data}, and it is \textbf{evaluated on biased data}.  Both of these can cause a significant problem in the system's ability to deliver user and business value and positive social impact.  Unfortunately, due to the unobservability of the construct feature and decision spaces, we do not know the precise structure of these biases, or the underlying true preferences or counterfactual responses.

Simulation can help with this.
We can simulate the entire data-generating process, from preference to observation.
Through this, we do \textbf{not} know if our simulation matches the process by which data in any actual system comes to be, but we do know the relationship between truth and observation in our data.
We can use such simulated data to study the distortions in recommender system behavior (and metrics of that behavior or performance) between what would be observed in an experiment with observable data, and what would be observed in an experiment with access to the actual underlying truth through an oracle.
One of my students has used this approach to measure biases in evaluation metrics that are induced by data missingness \citep{Tian2020-vs}, and \citet{Canamares2018-ki} employed a probabilistic model to better-understand popularity bias.

With simulation, not only can we create data sets that --- subject to certain assumptions --- provide both observations and their underlying truth, we can \textbf{change those assumptions} and re-run the simulation.
This allows us to study the sensitivity of our analysis to those assumptions; for example, if the bias in an evaluation metric is relatively stable across a range of plausible assumptions, that provides evidence that getting the data generating process exactly right is not so important and the metric may be reliable, but if it changes substantially with modest changes in assumptions, then we should treat results on that metric as highly tentative.

\section{Prospective Simulation: Studying Future Behavior}

Simulation also allows us to estimate possible future behavior and impact of the system and its users, under controlled and variable conditions, particularly as the system and users respond to each other.
Both \citet{Chaney2018-us} and \citet{Fleder2009-bi} have used simulation to study \emph{homogeneity} effects in recommender systems: to quantify the extent to which recommenders push users to consume the same items vs. distribute their attention across a wide range of diverse items.
This use of simulation has a wide range of applications, from homogeneity and popularity bias to filter bubbles to fairness concerns \citep{DAmour2020-zd} and many others regarding both a system's effectiveness and its impact on users, content creators, and society.

The key idea of many these simulations is to simulate the process of users consuming items, producing traces for training the algorithm, receiving recommendations, and consuming more items, possibly in response to those recommendations.
These models enable researchers to encode a wide range of assumptions into the user response models and study system performance and behavior under varying conditions.

\section{Recommender System Response Curves}

One of the major things that higher-degree simulation (anything above static data) affords in both of these, and other, scenarios is the ability to map out \textbf{response curves} for a recommender system or its surrounding experiments.
In our study of recommender system metric bias \citep{Tian2020-vs}, for example, we could extend the simulation to specifically model a variety of known degrees of popularity bias or of data sparsity, and estimate how the evaluation metric bias changes as a function of known changes in data biases.
We don't necessarily know the degree of bias that is present in real data, but if we can understand the evaluation process's response curve to that bias, it will produce knowledge that can be combined with future research that may provide a better idea of where in the curve any particular actual system lies.

We have a similar set of problems when working on counteracting potentially discriminatory biases in recommender systems.
We do not know, for example, what the distribution of author gender in book ratings and recommendations would be in an ideal world with no discriminatory factors affecting book production, reading, and recommendation \citep{Ekstrand2021-iu} (\citet{Mitchell2020-mt} identify this as ``the world as it can and should be'' in their taxonomy of biase sources); through simulation, we can quantify system behavior and response under a range of possible world-states and targets.
Existing and future research in a variety of fields will hopefully yield context to know what these respons cuves say about our existing world and book publishing ecosystem.

\section{Conclusion}

Simulation is a powerful tool for reckoning with uncertainty about what lies behind our data, or about how recommender systems may behave in the future under various conditions.
There is a lot of work to be done in order to understand how to develop, tune, and validate these simulations, but simulation has the promise to unlock types of research that are infeasible by any other means.

One of these is to to examine system behavior under specific, controlled conditions, and isolate the effect of particular user, item, or ecosystem dynamics on recommender behavior and user response.
Static data and actual applications differ in too many variables simultaneously to facilitate direct comparison that demonstrates the effect of specific features, but simulation allows us to create data sets or online responses that differ only in selected ways.
This will enable us to understand the behavior and effects of recommender systems and hypothetical human responses under a range of plausible and extreme conditions, and better understand when systems exhibit what behavior.
The ability to build systems that truly promote human flourishing and avoid harm depends on this analysis.

\begin{acks}
This paper based on work supported by the National Science Foundation under Grant No. IIS 17-51278.
\end{acks}

\bibliographystyle{ACM-Reference-Format}
\bibliography{multiverse}

\end{document}